\documentclass[fleqn,usenatbib]{mnras}
\usepackage[T1]{fontenc}
\usepackage{hyperref}
\usepackage{graphicx}	
\usepackage{amsmath}	
\usepackage{amssymb}	

\usepackage{newtxtext,newtxmath}
\usepackage[dvipsnames]{xcolor}
\usepackage{hyperref}
\graphicspath{{Figures/}}
\usepackage{float}
\usepackage{placeins}
\usepackage{lineno}

\newcommand{\apophis}{Apophis\ }
\newcommand{\Apophis}{Apophis\ }
\newcommand{\NA}{--}
\newcommand{\rev}[1]{\textcolor{black}{#1}}
\newcommand{\rvv}[1]{\textcolor{black}{#1}}
\newcommand{\rrv}[1]{\textcolor{black}{#1}}
\newcommand{\rrr}[1]{\textcolor{black}{#1}}

\title[2029 Observation Plan of Apophis]{Conditions for high-resolution bistatic radar observations of \apophis in 2029}

\author[Vallejo, Zuluaga \& Chaparro]{
Agustín Vallejo\thanks{E-mail: agustin.vallejo@udea.edu.co},
Jorge I. Zuluaga,
Germán Chaparro
\\
SEAP/FACom,  
Instituto de F\'{\i}sica - FCEN, Universidad de Antioquia
Calle 70 No. 52-21, Medell\'in, Colombia.\\
}

\date{Accepted XXX. Received YYY; in original form ZZZ}

\pubyear{2021}

\begin{document}
\label{firstpage}
\pagerange{\pageref{firstpage}--\pageref{lastpage}}
\maketitle

\begin{abstract}
On \rev{April 13, 2029}, asteroid \Apophis will pass within {six} Earth radii ($\sim$31000 km \rvv{above the} surface), in the closest approach of this asteroid in recorded history. This event provides unique scientific opportunities to study the asteroid, its orbit, and surface characteristics at an exceptionally close distance.
\rvv{In this paper, we} perform a \rev{novel} synthetic geometrical, geographical and temporal analysis of the conditions under which the asteroid can be observed from Earth, with a particular emphasis on the conditions and scientific opportunities for bistatic radar observations, the most feasible radar technique applicable during such a close approach.
For this purpose, we compile a list of present and future radio observatories \rvv{or radio facilities} around the globe which could participate in bistatic radar observation campaigns during the close approach of Apophis.  We estimate signal-to-noise ratios, apparent sky rotation, surface coverage and other observing conditions. 
We find that a global collaboration of observatories across Australia, Africa, Europe and America will produce high-resolution delay-Doppler radar images with signal-to-noise ratios above $10^8$, while covering $\sim$85\% of the asteroid surface. Moreover, if properly coordinated, the extreme approach of the asteroid might allow for radio amateur detection of the signals sent by large radio observatories and citizen science projects could then be organized. We also find that for visual observations, the Canary Islands will offer the best observing conditions during the closest approach, both for professionals as well as for amateurs.  The apparent size of \Apophis will be 2-3 times larger than typical seeing, allowing for resolved images of the surface.

\end{abstract}


\section{Introduction}

Previous observations of the trajectory of asteroid 99942 \Apophis have shown that on April 13, 2029, it will pass within six Earth radii from our planet. Since its discovery in 2004, the probability of impact has been reduced from 2.7\% to practically nil \citep{Chesley2005, Wlodarczyk2013}. In particular, recent radar observations of Apophis have mostly ruled out this risk \citep{brozovic2022}. However, Apophis still provides unique opportunities in terms of asteroid studies, which are usually limited to very distant and faint objects, and relatively small ones passing close-by (in the tens of meters range), theoretical modeling, or morphology estimations \citep{Herique2020}. This time, the asteroid will come to us. 

\cite{Pravec2014} and others have revealed Apophis to be an elongated object of $340\pm40$ m on its semimajor axis, axial anisotropy $a/b$ of 1.44, and rotational period of $\sim$30.56 h. Observations of Apophis using Goldstone and Arecibo by \cite{Brozovic2018a} have given further clues to its particular shape, albedo, and other physical features such as its radar cross-section. While they report no Yarkovsky acceleration, \cite{Farnocchia2020} did detect a semimajor axis drift of about $\sim$170 m per year. These observations were all prior to the close encounter of March 2021 at 0.12 au \citep{Chesley2005}, which will probably \rev{settle the issue of the role} of the Yarkovsky effect.

It has been predicted that for the 2029 fly-by, Apophis' spin will significantly change from its current state \citep{Scheeres2005,Souchay2018}, while its shape will remain mostly unaffected by tidal forces \citep{Yu2014}. Regarding solar radiation pressure, it has been shown that its effect is minimal on the orbit of Apophis \citep{Giorgini2008, Zizka2011}. Therefore, taking into account known non-gravitational effects, the ephemeris can be calculated with an uncertainty smaller than a few kilometers\footnote{NASA JPL Horizon 
\href{https://go.nasa.gov/3D7kk9W}{data} retrieved at \today{}} which gives us a unique opportunity to prepare collaborative strategies for observing the asteroid well in advance of its fly-by. 

Although \rev{optically resolved} observations of Apophis will be possible during its closest approach (see \autoref{sec:approach}), it will be visible mostly from the day-side of the planet, thus preventing high-quality observations. Fortunately, Apophis will fly \rev{over some of the most important radar facilities and radio telescopes on Earth, such as Goldstone, Very Large Array (VLA), next-generation VLA (ngVLA), Haystack Ultrawideband Satellite Imaging Radar (HUSIR), Green Bank (which is currently being fitted with a new, 100 kW transmitter, see \citealt{decadal})}, and also China’s Five-hundred-meter Aperture Spherical radio Telescope (FAST). 

A \rvv{combination} of bistatic radar observations of this asteroid from multiple radio facilities will allow for a detailed map of the asteroid down to the highest possible range resolution, which is currently 1.875 m/px using the Goldstone Deep Space Station-13 antenna (DSS-13, \citealt{Naidu_2016}). \rrv{In fact, the achievable range resolution could potentially be close to two orders of magnitude better if using satellite imaging radar facilities which have not yet seen an asteroid pass within their range}. This would allow for a detailed study of asteroid properties for which radar astronomy is well suited for: high-cadence measurements of its spin state, surface dust properties, and possible tidal effects caused by its close approach such as asteroid-quakes. Additionally, due to the short round-trip time at the closest approach, bistatic radar has the best potential for achieving a high delay-Doppler frequency resolution with an integration time that is not restricted by the TX-RX switch time of monostatic observations.

The case for planetary radar observations has been strongly supported by \cite{Brozovic2018b} and \cite{Virki2020}. In particular, \cite{Herique2020} has advocated for the use of this technique for monitoring observations of Apophis. In the case of the Apophis 2029 fly-by and other objects experiencing close encounters with Earth, radar signals have round-trip travel times that are too short for monostatic radar observations. This means that radar observation campaigns for the 2029 Apophis fly-by have to be mostly bistatic, i.e. where the transmitting and receiving antennas are not the same.

For this reason we propose a sequential bistatic radar observation plan during the day of closest approach of Apophis, calculating the expected radar parameters such as the apparent asteroid sky rotation, \rvv{signal-to-noise ratio (SNR) per run}, etc., based on its expected trajectory, spin state, and observation conditions to an unprecedented level of detail. We identify the best pairings of radar facilities and suitable radiotelescopes in terms of the best achievable SNR and range resolution.

\rvv{This paper is organized as follows:  In \autoref{sec:approach} we describe the details of the 2029 fly-by: the approaching conditions of \apophis around Earth, its magnitude, apparent size, distances, etc. A detailed analysis of the conditions for making paired radar observations using radio antennas around the world is described in \autoref{sec:coordinated}. The technique of bistatic radar observations is described in \autoref{sec:bistatic}. In \autoref{sec:rotation} we discuss on the impact the rotation of \apophis would have on radar observations. Along the paper we emphasize on how this flight represents a potential citizen science opportunity, and how even amateur radio astronomers can take part in this important event.}

\section{Approaching \rvv{window} and visual observations}\label{sec:approach}

Radar observations of \apophis will start to be feasible from early March up to June 2029 (a couple of months before and after its closest approach) when the asteroid will come at a distance of $\sim2000\;R_\oplus$ ($\sim0.009$ AU or $\sim33$ times the Earth-Moon distance) from our planet \citep{Brozovic2020}. However, at its closest approach Apophis will be at a distance of near $5\;R_\oplus$ from Earth. Around 4 hours before its closest approach (April 13, 21:50 UTC\footnote{All the information about Apophis trajectory used in this work was obtained from JPL’s Horizons web service, and processed in Python with {\tt Astropy} and {\tt Cartopy} libraries.}), the apparent rotation rate of the asteroid (see \autoref{sec:rotation}) will increase, while the distance to Earth will decrease at a fast pace, increasing significantly the signal-to-noise ratio for bistatic radar observations (see \autoref{sec:bistatic}). Coincidentally, around 8 hours before closest approach, when the asteroid is at a distance of ${\sim}30 R_\oplus$, its visual magnitude will reach the human eye limit. For these two reasons, we have chosen the period of time when the asteroid is inside $30 R_\oplus$, that covers from April 13, 14:00 UTC to April 14, 06:00 UTC, as the time-window to focus our analysis (we will call it the {\it approaching window }).

\begin{figure}
    \centering
    \includegraphics[width=\linewidth]{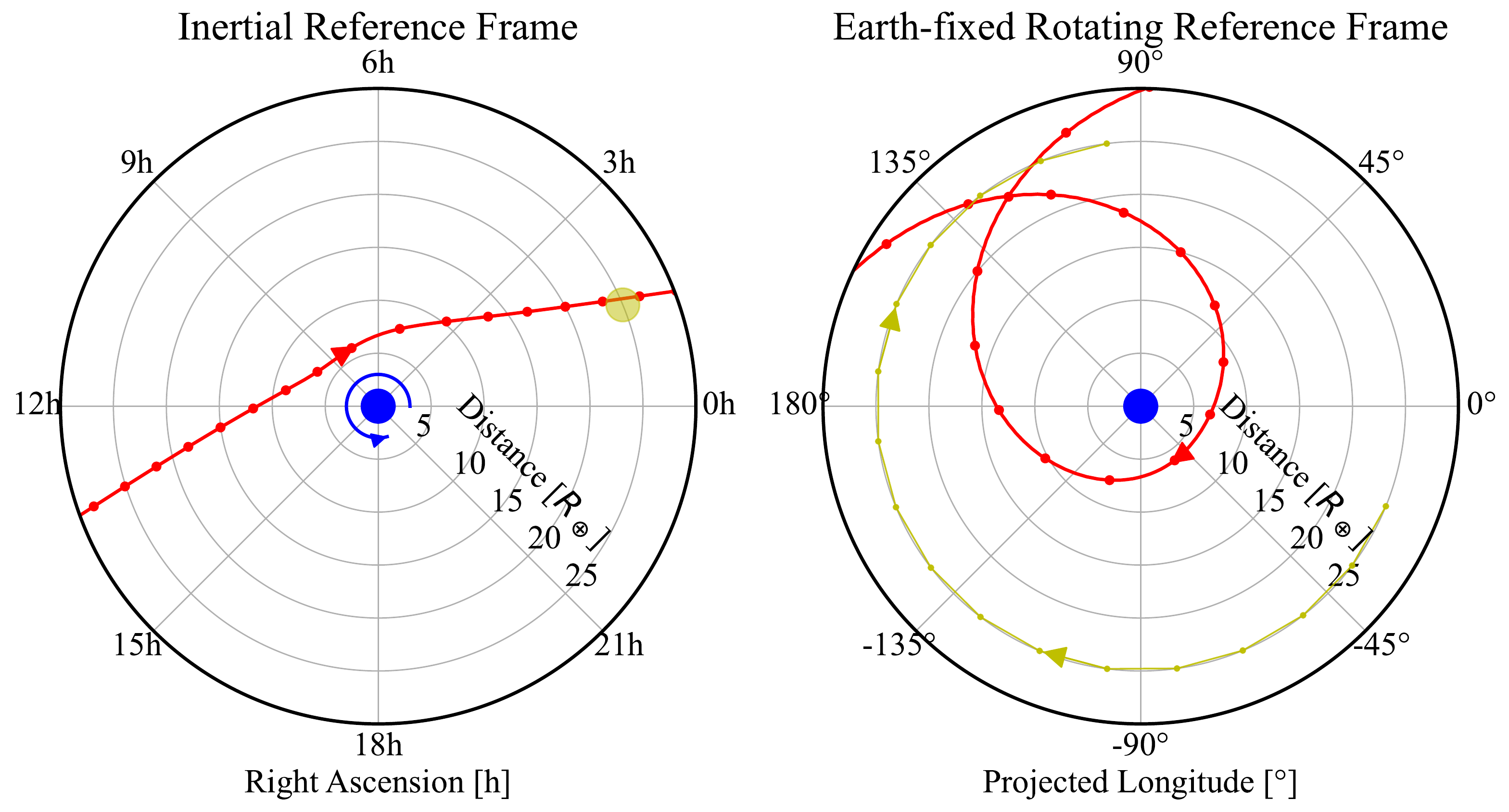}
    \caption{Distance of asteroid \Apophis to Earth's center in earth radii during the hours around the closest approach. \rev{Left panel} shows the right ascension of the asteroid and the geocentric distance (projection of an hyperbola onto the equatorial plane). \rev{Right panel} shows the projected geographical longitude of the Apophis as seen from a reference frame fixed on the rotating Earth. At its closest approach the asteroid will fly-by over the atlantic (lon. $43.4^{\circ}W$, lat. $28.9^{\circ}N$) at a distance of $6.0 R_\oplus$ around 18:50 local time (21:50 UT). \rev{The yellow circle represents the Sun's position in each frame of reference. The points in the paths are 1h marks.} }
    \label{fig:distance}
\end{figure}

In \autoref{fig:distance} we show the path of the asteroid in the approaching window, both in an \rvv{Earth-centered} reference frame (fixed on the stars) and in a rotating reference frame (fixed on the Earth). As we see there, \Apophis closest approach will occur at $6.0\;R_\oplus$ with respect to Earth's center, precisely over a point on the Earth' surface with longitude (lon.) $43.4^{\circ}W$ and latitude (lat.) $28.9^{\circ}N$ (a point on the Atlantic ocean, northwest of Canary Islands). 

\begin{figure}
    \centering
    \includegraphics[width=\linewidth]{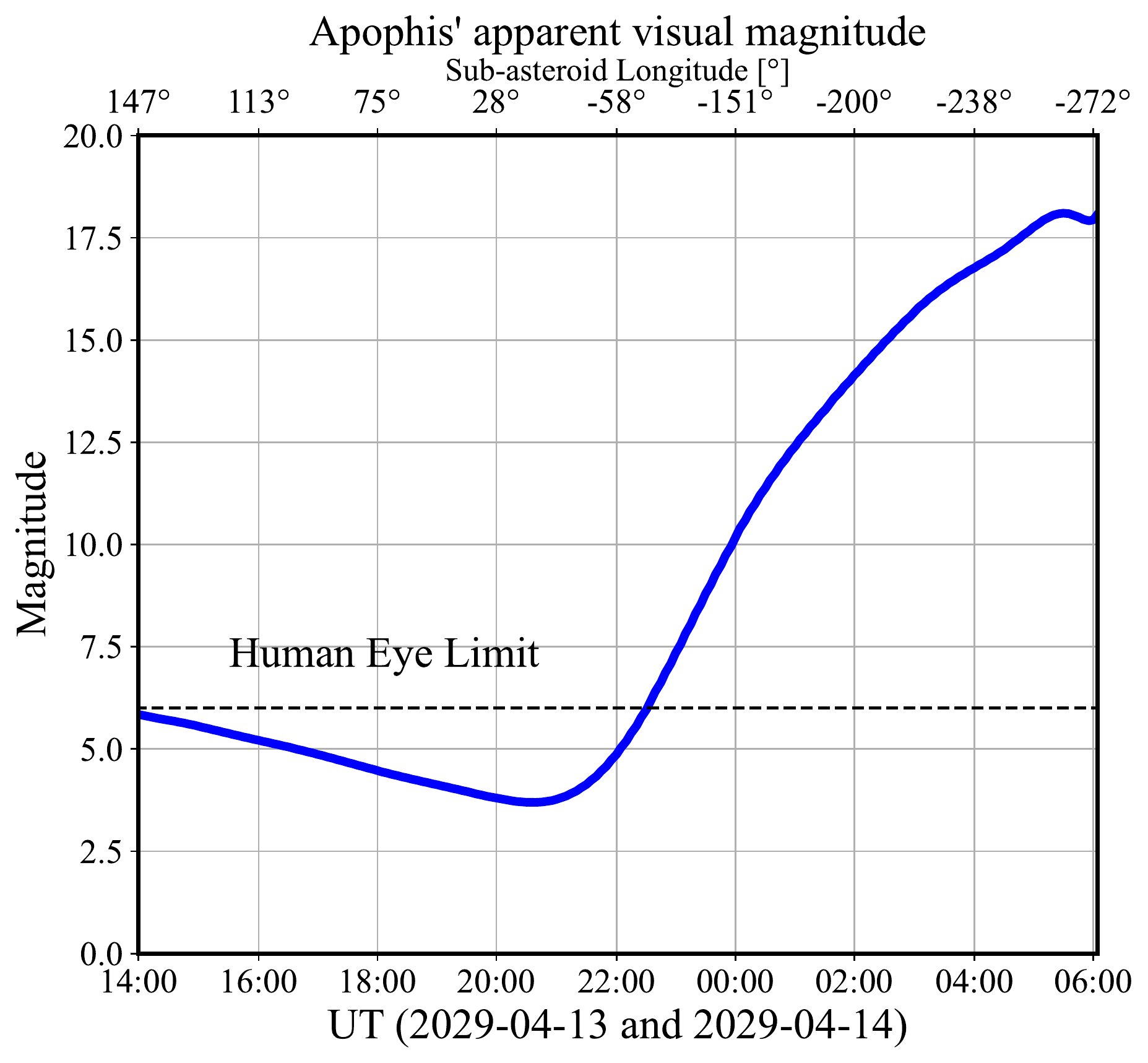}
    \caption{Apparent magnitude of \Apophis on the hours around closest approach (21:45 UTC). As it gets closer, it will become brighter, reaching a minimum visual magnitude of 3.7 at 20:35 UTC. Right after closest approach the asteroid will go into the illuminated side of Earth. Alongside the top x axis, the projected asteroid's latitude is described \rev{for each time to determine at which meridian the asteroid will be visible.}}
    \label{fig:mag}
\end{figure}

The asteroid will be visible to the naked eye mostly over Europe, Africa and western Asia from the beginning of the approaching window and up to 2 hours after the closest approach (see \autoref{fig:mag}). Then, the asteroid will experience a rapid decrease in apparent brightness due to phase variations and the fact that it will be flying over the sun-illuminated side of the Earth. Only radar observations will still be feasible then.

\begin{figure*}
    \centering
    \includegraphics[width=\textwidth]{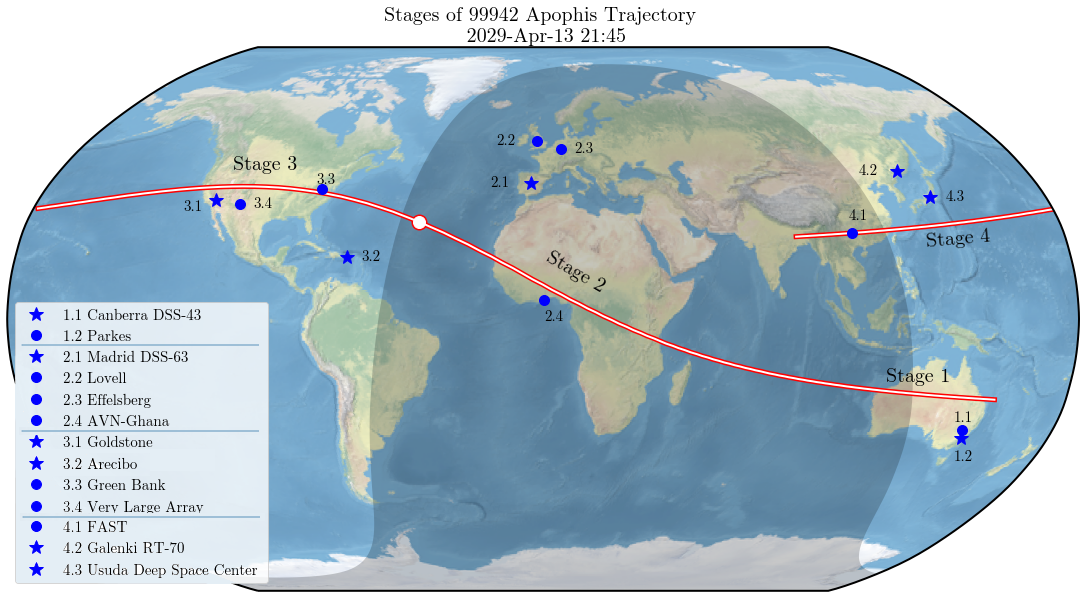}
    \caption{Projection of Apophis' path over Earth's surface along with major radio telescopes in the world (stars are transmitting facilities whereas blue circles are receiving facilities) across its path. The red circle is the point of closest approach over lon. $43.4^{\circ}W$, lat. $28.9^{\circ}N$ and distance of $6.0 R_\oplus$. The trajectory was divided in four stages based on geographical sub-asteroid point: Australia, Africa (and Europe), USA, and Asia. (See \autoref{sec:coordinated})}
    \label{fig:pathWorld}
\end{figure*}

In \autoref{fig:pathWorld} we show the projection of \Apophis' path over the surface of the Earth during the approaching window (i.e. when the asteroid is closer than $30 R_\oplus$). We also show the sub-asteroid coordinates, the distance and apparent magnitude of the asteroid on 2 hour intervals in \autoref{table:coords}. As we see in the map, the projected path during the approaching window  will start over Australia and then it will move rapidly towards southeast Africa and cross the Atlantic ocean towards America.  At the end of the approaching window the asteroid will be flying over northeast Asia.

\begin{table}
\begin{tabular}{lcccc}
\hline\hline
Time [UTC] & Lon. [deg] & Lat. [deg] & Dist. [$R_\oplus$] & V mag. \\ \hline
2029-Apr-13 14:00          & 147.4          & -23.5         & 28.5         & 5.8          \\
2029-Apr-13 16:00          & 113.1          & -21.0         & 21.7         & 5.2          \\
2029-Apr-13 18:00          & 75.5           & -15.9         & 15.0         & 4.5          \\
2029-Apr-13 20:00          & 27.9           & -2.3          & 8.8          & 3.8          \\
\textbf{2029-Apr-13 21:45} & \textbf{-43.4} & \textbf{28.9} & \textbf{6.0} & \textbf{4.4} \\
2029-Apr-14 00:00          & -151.3         & 36.3          & 10.3         & 10.0         \\
2029-Apr-14 02:00          & 159.8          & 30.1          & 16.7         & 14.0         \\
2029-Apr-14 04:00          & 122.4          & 26.6          & 23.4         & 17.0         \\
2029-Apr-14 06:00          & 88.4           & 24.5          & 30.2         & 18.0   \\\hline\hline     
\end{tabular}
\caption{\Apophis projected position, distance and (geocentric) apparent magnitude on 2h intervals during the selected time-window.\label{table:coords}}
\end{table}

\begin{figure}
    \centering
    \includegraphics[width=\linewidth]{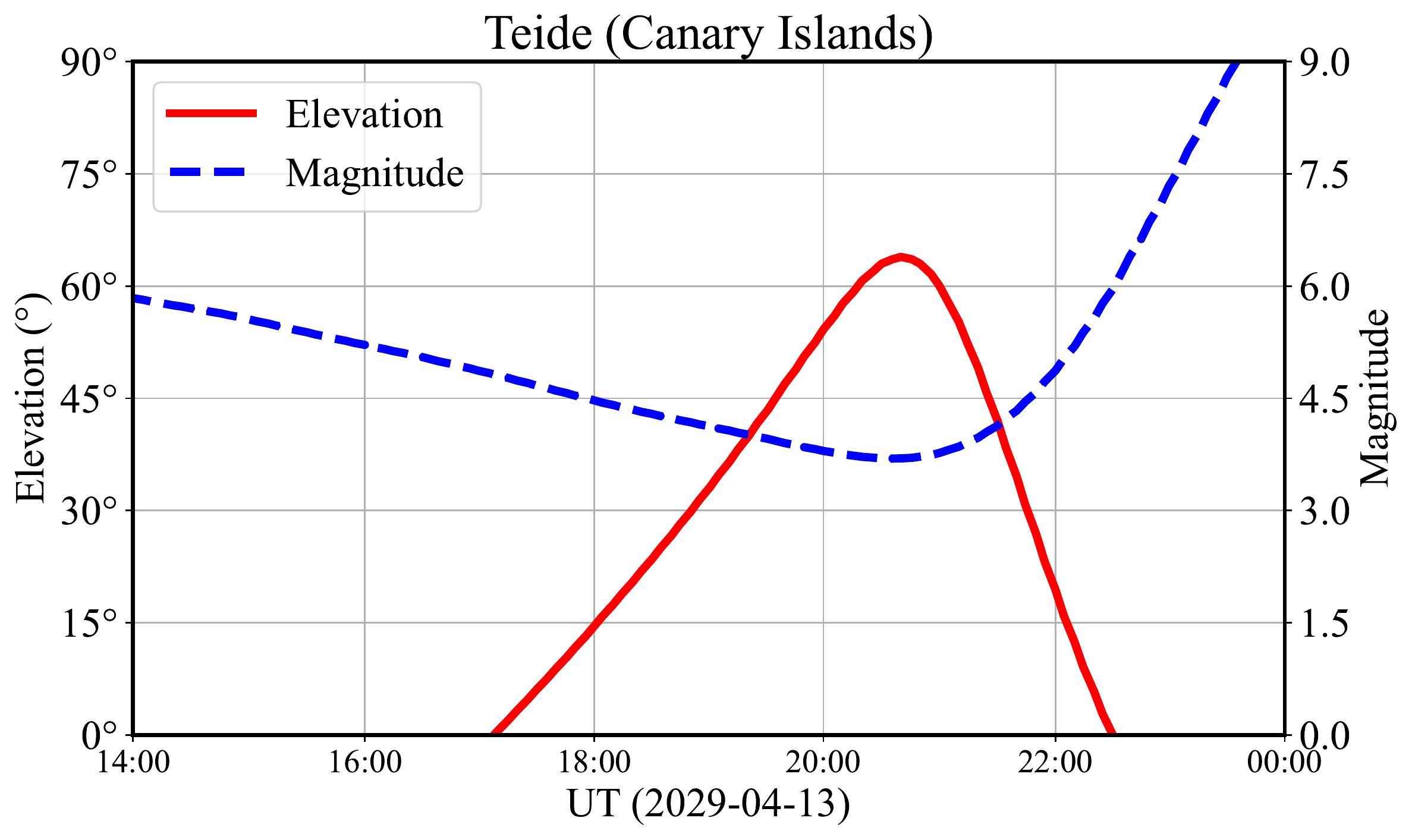}
    \caption{\rev{Elevation and apparent} magnitude of \Apophis during its closest approach to Earth in April 13, 2029, as seen in Teide (Canary Islands).}
    \label{fig:elevationTeide}
\end{figure}

One of the best suited \rrv{close-approach} observation regions on Earth are the Canary Islands.  In \autoref{fig:elevationTeide} we show the elevation and apparent magnitude 
of \apophis as seen from the top of mount Teide in Tenerife\footnote{We have created a website that provides the visual observing conditions for any location on Earth \url{https://agustinvallejo.github.io/apophis.html}} (an ideal place for professional as well as amateur observations).  Using the size estimation of \cite{Brozovic2018a} of $340\pm40$ m the maximum apparent size of \apophis will be of $1.82\pm0.21''$ which is 2-3 times larger than the \rrv{local} seeing \footnote{For typical observing conditions in the El Teide or Roque de los Muchachos sites see \url{https://www.ing.iac.es/astronomy/observing/conditions/enofolleto.pdf}, page 6 \rvv{(last accessed \today{})}.}.  Although the optical observing window is much \rrv{shorter than for radar observations, optical observations will help in constraining the shape in addition to the radar data.}. \autoref{table:opticalObs} also shows a list of significant optical telescopes that are \rrv{well-located for observing the close approach}.
\begin{table*}
\label{table:opticalObs}
\begin{tabular}{llrrr}
\hline\hline
\textbf{Name} & \textbf{Location} & \textbf{Longitude (°)} & \textbf{Latitude (°)} & \textbf{Diameter (m)} \\
\hline\hline
Devasthal & Nainital district, Kumaon division, Uttarakhand, India & 79.684 & 29.36 & 3.60 \\
BTA-6 & Caucasus Mountains, Russia & 41.44 & 43.65 & 6.00 \\
SALT & Great Karoo, Sutherland, South Africa & 20.81 & -32.38 & 10.50 \\
Calar Alto Observatory & Almería, Spain & -2.55 & 37.22 & 3.50 \\
William Herschel & Santa Cruz de Tenerife, Canary Islands, Spain & -17.88 & 28.76 & 4.20 \\
Gran Telescopio de Canarias & La Palma, Canary Islands, Spain & -17.89 & 28.76 & 10.40 \\
Galileo National Telescope & La Palma, Canary Islands, Spain & -17.89 & 28.76 & 3.58 \\ \hline\hline
\end{tabular}
\caption{\rev{Some of the optical telescopes that could make observations of the asteroid in visible light, sorted by longitude.}}
\end{table*}
\section{Coordinated radar observations} \label{sec:coordinated}

One hour after the closest approach, \Apophis will be over North America, and although it will be daytime, the asteroid will be flying overhead the area of the Goldstone complex and the Green Bank radio observatory. This will be one of the ideal moments for {\it bistatic radar observations} (see \autoref{sec:bistatic}) as these radio facilities have large antennas, significant experience with planetary radar astronomy and a clear opportunity for observing the asteroid (at that time \apophis will still be at ${\sim}7\;R_\oplus$ from Earth's center, or ${\sim}6\;R_\oplus$ from the telescopes). 

As a matter of fact (as shown in \autoref{fig:pathWorld}), most of the largest radio facilities on Earth with transmission capabilities (stars in the map) will be able to \rvv{participate in} radar observations of the asteroid.  However, during the approaching window, the close proximity of the asteroid and the fast movement of it through the sky will make {\it monostatic radar observations} \rrv{hard to achieve}: if a single facility has transmission (TX) as well as receiving (RX) capabilities, it will be very difficult for it to switch \rrv{modes} in time \citep{Virki2020}. However, although most radio telescope facilities do not have TX capabilities, a significant number have significant RX sensitivity. This is why bistatic radar observations around the globe would offer the best opportunities to observe the closest approach of Apophis.

The next significant moment of the fly-by will occur 7 hours after the closest approach, when the asteroid passes over NE Asia. In particular, it will fly very close to the zenith of the FAST radio telescope.  Since FAST is not fully steerable, the fact that the receding, still close \Apophis will fly almost vertically over FAST is truly fortunate.

\begin{figure*}
    \centering
    \includegraphics[width=\textwidth]{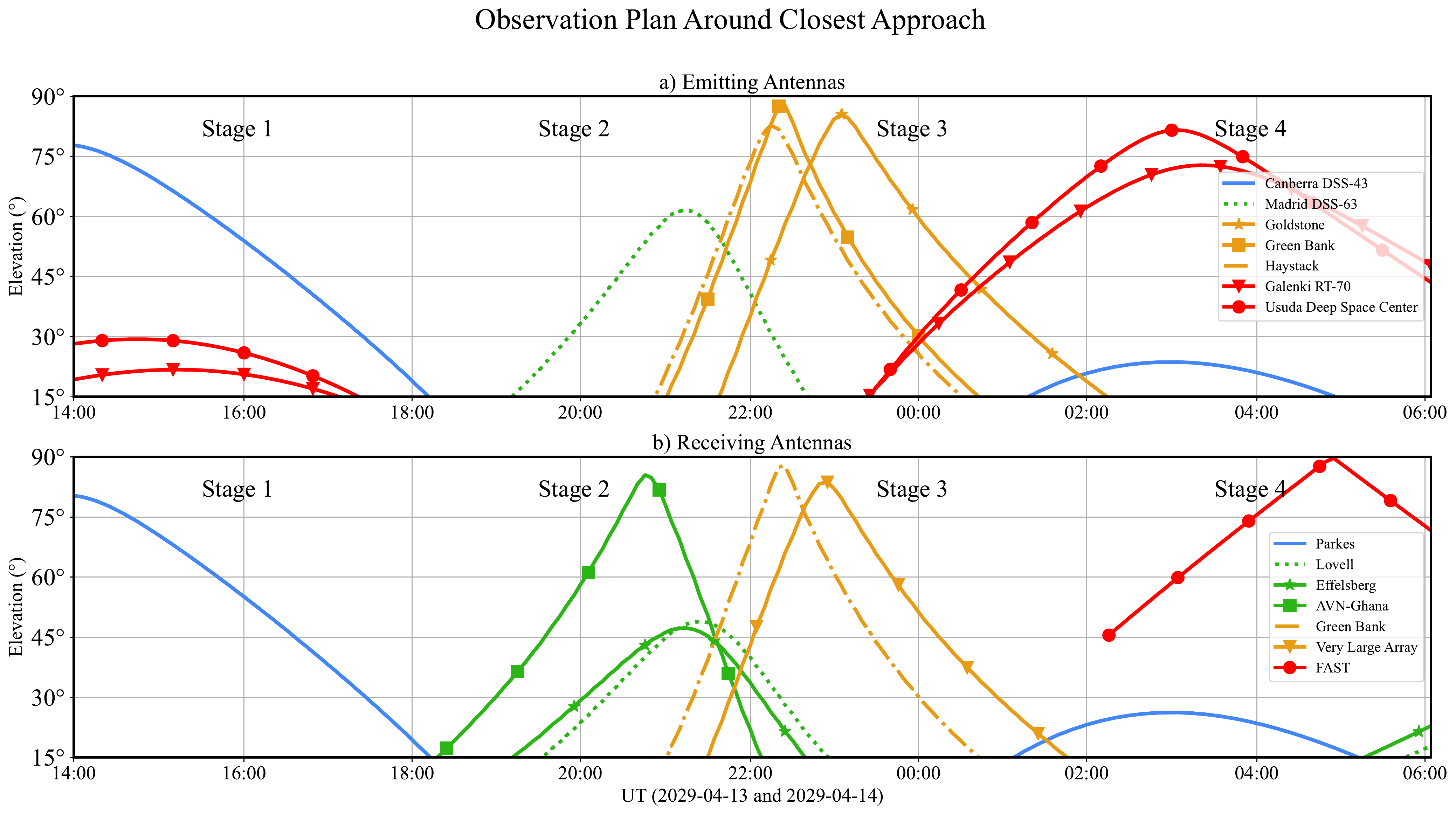}
    \caption{Proposed observation plan of Apophis closest approach, which will get almost complete coverage of the fly-by. a) the expected elevations of the asteroid as seen from the \rvv{transmitting} facilities. b) the elevations from the receiving end. We suggest the 4-stage plan of the transit: the first one being Canberra DSS-43 paired with Parkes, as they have optimal view of the asteroid, then with the African and European dishes (Madrid DSS-63 with AVN-Ghana, Lovell or Effelsberg). This stage will see the asteroid at its closest. Third stage will be Goldstone with Green Bank or VLA, which will also get the closest point. Fourth stage is FAST with Asian or Russian antennas as transmitters. \rvv{Note that the y-axis goes from 15° to 90°, because the asteroid being over the horizon does not guarantee a successful observation, all antennas have a minimum elevation, most of them around 15°. FAST can only point above $\sim$45°, depending on the mode of observation.} }
    \label{fig:pairs}
\end{figure*}

In order to organize a global radar observational effort, we propose to split the approaching window in four {\it stages}. At each stage we identify the best pairings of radar facilities and observatories for performing bistatic radar observations.  In \autoref{tab:observatories} we show the location, and the RX and TX properties of the radio observatories included in this work.  For performing the pairing analysis, we will refer to \autoref{fig:pairs}, where we show the elevation of the asteroid as a function of time for all the observatories listed in \autoref{tab:observatories}. Additionally, in \autoref{table:crossMatching} we pair some of the most important radio facilities to show which ones would be available at the same time for bistatic observations.

\rvv{Furthermore, simultaneous bistatic observations (e.g. when DSS-13 and DSS-14 transmit at the same time) will let us explore the asteroid beyond its shape from the radar albedo and polarization at different wavelengths. This will help constrain the surface composition and regolith grain size distribution \citep{albedo}. }

\rrv{In the following sections we will describe the observation plan along with the achievable per run SNR for significant bistatic radar pairs in each region. Here each \emph{run} refers to the minimum integration time needed for achieving full Doppler frequency resolution (see \autoref{eq:trun}).}

\begin{table*}
\begin{tabular}{llcccccc}
\hline\hline
Stage & Name                    & Lon. [deg] & Lat. [deg] & D [m] & $P_\mathrm{tx}$ [kW] & $F_\mathrm{tx}$ [GHz] ($\lambda$ [cm]) & $F_\mathrm{rx}$ [GHz] \\ \hline\hline
1.1   & Canberra DSS-43         & 148.98    & -35.40   & 70      & 80 & 7.2 (4.2) & \NA  \\
1.2   & Parkes                  & 148.26    & -33.00   & 64      & \NA & \NA  & 0.8-22  \\\hline
2.1  & Madrid DSS-63           & -4.25     & 40.43    & 70      & 20  & 7.2 (4.2)  & \NA\\
2.2   & Lovell                  & -2.31     & 53.24    & 76      & \NA & \NA  & 0.4-6   \\
2.3   & Effelsberg              & 6.88      & 50.52    & 100     & \NA & \NA  & 0.3-95  \\
2.4   & AVN-Ghana               & 0.31      & 5.74     & 32      & \NA & \NA & 4-8   \\
2.5    & TIRA (*)                    & 7.12      & 50.66    & 34      & 1000 & 22.5 (1.33)  & \NA\\
2.6    & Yevpatoria RT-70        & 33.18    & 45.18    & 70       & 100  & 5 (6)  & \NA \\
2.7    & Sardinia                & 9.25      & 39.48    & 64      & \NA & \NA & 0.3-116  \\\hline
3.1   & Goldstone DSS-13        & -116.89   & 35.43    & 34      & 80 & 7.2 (4.2) & \NA   \\
3.1   & Goldstone DSS-14       & -116.89   & 35.43    & 70      & 450 & 8.6 (3.5)  & \NA \\
3.2   & Arecibo (defunct)                & -66.75    & 18.34    & 305     & 2000 & 2.4 (13) & \NA  \\
3.3   & Green Bank \rrr{(**)}              & -79.84    & 38.43    & 100     & 100 & 35 (0.85)  & 0.3-116  \\
3.4   & Very Large Array        & -107.62   & 34.08    & 25 ea.     & \NA & \NA  & 0.06-50 
\\
3.5   & Haystack (HUSIR) (*)       & -71.42   & 42.68    & 36.6     & 250 & 10 (3)   & \NA  
\\\hline
4.1   & Galenki RT-70           & 131.76    & 44.03    & 70      & 80  & 5 (6) & \NA  \\
4.2   & Usuda Deep Space Center & 138.36    & 36.13    & 64      & 20  &  8 (3.7)  & \NA\\
4.3   & FAST                    & 106.86    & 25.65    & 500     & \NA & \NA & 0.07-3    \\ \hline\hline  
\end{tabular}
\caption{Relevant radio facilities in the asteroid's path. Those with listed power are the transmitting facilities considered here. (*) The average power for pulsed radars is an order of magnitude lower than the peak power referenced here. \rrr{(**) The quoted $F_{\mathrm{tx}}$ value of 35 GHz for the GBT is a preliminary value, and could change in the future.}\label{tab:observatories}}
\end{table*}

\begin{table*}
\label{table:crossMatching}
\begin{tabular}{lllllll}
\hline \hline
\rvv{RX / TX} & Canberra DSS-43 & Madrid DSS-63 & Goldstone & Green Bank & Galenki RT-70 & Usuda \\ \hline
Parkes & X &  &  &  & X & X \\ \hline
Lovell &  & X & X & X &  &  \\ \hline
Effelsberg &  & X & X & X &  &  \\ \hline
AVN-Ghana &  & X &  & X &  &  \\ \hline
VLA &  &  & X & X &  &  \\ \hline
Green Bank &  &  & X &  &  &  \\ \hline
FAST & &  &  &  & X & X \\ \hline \hline
\end{tabular}
\caption{Available pairing of some of the proposed radio observatories based on elevation and available window of observation. \rvv{In the first} column are the receivers and on the first row are the possible transmitters.}
\end{table*}

\subsection{Stage 1. Australia}
At the beginning of the approaching window, \Apophis will fly directly over Australian territory.  This moment will offer a good opportunity to transmit from Canberra's DSS-43 antenna. \rrr{This is} one of the largest steerable antennas in the hemisphere (70 m diameter) with a recently installed 100 kW transmitter. An S-band 400 kW\footnote{See the DSN 101 10-m Subnet Telecommunications Interfaces report \url{https://deepspace.jpl.nasa.gov/dsndocs/810-005/101/101E.pdf} \rvv{(last accessed \today{})}.} and an X-band 80 kW\footnote{See the DSN 301 Coverage and Geometry Report \url{https://deepspace.jpl.nasa.gov/dsndocs/810-005/301/301N.pdf} \rvv{(last accessed \today{})}.} transmitter are also available. \rvv{The Parkes radio telescope will also be under the asteroid's path, as well as any of the Australia Telescope Compact Array (ATCA) elements, which we do not consider in our calculations but could plausibly be used as RX}. 

Apophis will be over the horizon with respect to DSS-43 and Parkes from 9:00 UTC to 19:00 UTC, and the distance will decrease from $40 R_\oplus$ to $10 R_\oplus$ during this time window. The flight path will allow for an elevation of almost 80°.

Last, but not least, this stage could also be ideal for visual observations of \apophis, as it will be close to midnight in western Australia.  Because at this stage the visual magnitude of the asteroid will be very close to the human eye limit, most telescopes will be able to observe the approaching object.

\subsection{Stage 2. Africa and Europe}
The asteroid will continue its course westward and fly over Africa towards the Canary Islands. In this stage, the best antenna for transmission will be Madrid's DSS-63. Similar to the one in Canberra, DSS-63 has a 70m diameter and 20 kW of available TX power.  At apogee, DSS-63 will see the asteroid at 60$^{\circ}$ of elevation (see \autoref{fig:pairs}).

At the reception end, there are a few alternatives in Europe as well as in Africa.  In Europe there are the Effelsberg (100 m) and Lovell (76 m) radio telescopes, \rvv{both of which will} see a maximum elevation of about 45$^{\circ}$. \rvv{In Africa, for example, the Ghana Kuntunse radio telescope (32 m) will see the asteroid nearly in the zenith.} In the case that the Square Kilometer Array is operational in 2029 it could plausibly participate in these observations on the RX end. 

Other radio telescope arrays such as any of the European VLBI Network (EVN) antennas could also be used for this observation. For this, either the Madrid antenna, the Tracking and Imaging Radar (TIRA) in Germany or the Yevpatoria RT-70 antenna \citep{yevpatoria} could be used for transmission to increase the number of antenna pairs for bistatic observations.

This stage spans from 18:00 UTC to 23:00 UTC. During this time the asteroid will be fully visible for countries in Europe and Africa, even with the naked eye (see \autoref{fig:mag}). The geocentric distance during this stage, will change from $15\;R_\oplus$ to $6\;R_\oplus$ at 21:45 UTC and finally $7 R_\oplus$ at the end of this stage.

\subsection{Stage 3. Americas}

Just after the closest approach, the asteroid will continue its path over North America. Unfortunately (for visual observers) it is going to be daytime in this hemisphere and no visual observations will be attainable. However, the asteroid will fly right over important American radio telescopes. The Goldstone complex is an ideal facility for transmitting radar signals to the \rvv{asteroid surface, with its DSS-13 and DSS-14 antennas (TX frequency: 7190 MHz and 8560 MHz, respectively).  Previous radar observations have also used DSS-14 for mapping the shape and surface characteristics of Apophis at 18.75 m/px, since its 70 m dish and 500 kW transmitting power allowed for an SNR of $\sim23$/day \citep{Brozovic2018a}. However, for the 2029 approach the expected per run SNR for the smaller TX-capable antenna (DSS-13, 34 m dish) will be of the order of $10^7$ (see  \autoref{fig:SNR}). Thus, DSS-13 with its planned 7190 MHz transmitter, with an 80 MHz decoder bandwidth  \cite{brozovic2022} will yield a range resolution of 1.875 m/px (see \autoref{tab:observatories}) .}

The Green Bank Telescope (GBT, $D=$100 m) is also currently being fitted with TX capabilities \citep{decadal} with a 100 kW transmitter at 35 GHz planned to be installed before the Apophis approach.  Any of the VLA or VLBA dishes \rvv{would} be able to act as RX at the planned TX frequencies mentioned above.

We should note that sub-m range resolution could be achieved \rrv{when using Haystack Ultrawideband Satellite Imaging Radar, HUSIR} \citep{husir} reaching an instantaneous\footnote{For pulsed radars the subsecond (e.g., 1/60 Hz) duty cycle duration might be more appropriate, but at this point we only quote this as a representative SNR value.} (1 second) SNR of up to $\sim\times10^7$ for monostatic observations, with a 1024 MHz decoder bandwidth allowing for a range resolution of 25 cm/px at X-band (10 GHz), and with a $T_\text{sys}=100$ K. The average SNR will be an order of magnitude lower due to HUSIR being a pulsed radar facility instead of continuous-wave. HUSIR is also equipped with a W-band (96 GHz, 0.3 cm) transmitter with an 8 GHz effective bandwidth, allowing for a range resolution in the order of 3 cm/px \citep{widebandrf,husirrange,husir96ghz}. It should be also be noted that although this instrument is capable of satellite imaging, its asteroid imaging capabilities have not yet been fully tested at this resolution.

To match the best possible resolution, X-band observations require a rotation of at least 3.44$^\circ$ over the integration time \citep{widebandrf}, which for the case of Apophis is of the order of $\sim20$ min. W-band observations would require 8x more integration time.

Although HUSIR is capable of monostatic radar observations at very close distances, it is limited to observing anything within 50,000 km (for X-band) and 5,000 km for W-band. Therefore, we propose using GBT as RX in conjunction with HUSIR (as TX) for bistatic observations, although keeping in mind that the GBT dish has a very low aperture efficiency at W-band. Also, the W-band suffers a lot from atmospheric absorption, so the system temperature should be about an order of magnitude more than \rrr{the usual 25 K}. However, the achievable per-run SNR could be up to $\sim10^8$, given $T_\text{sys}=25$ K for bistatic observations, \rrr{or $\sim10^7$, for $T_\text{sys}=250$ K.}

Another ideal transmitting antenna at this stage would have been the Arecibo Telescope, whose location is very close to the flight path of the asteroid. It will be a great loss if a similar observatory were not to be built at the same place in the following years, although we are not aware of any such plan.  

\subsection{Stage 4. East Asia}
As the asteroid is finally leaving Earth's proximity on the morning of April 14th, it will fly over the 500 m FAST radio telescope. One candidate for transmitting is the Japanese Usuda radio station (64 m diameter and 20 kW of power). Another plausible TX facility is the 70 m Galenki (Ussuriysk), Russia antenna with 80 kW of transmitting power. \rvv{It should be noted FAST is limited in terms of frequency capabilities, as it can currently only observe up to 3 GHz (S-band), whereas the Russian antennas do not transmit at S-band but at 5 GHz. The Apophis flyby can thus be an important scientific case for expansion and improvement of these facilities. Therefore, this stage is only feasible if those upgrades are implemented and completed before the flyby.}

\section{Sensitivities and resolution} \label{sec:bistatic}

Asteroid radar observation is a technique that depends on many geometrical and physical factors \citep{ostro1993,hudson1994,ostro2002}. Due to the high level of control with respect to technical aspects, this technique can be considered to be no more complicated than obtaining visual observations, as the illumination is always well constrained and the corresponding phase angle is near $0^\circ$ even for bistatic observations. In this section, we will use the ephemeris of the 2029 \Apophis approach to estimate relevant quantities involved in bistatic radar observations of the asteroid. Our main goal here is to identify the scientific opportunities of a coordinated global collaboration for mapping a significant fraction of the surface of \Apophis at unprecedented SNR and range resolution.

The delay-Doppler technique is often used (see e.g. \citealt{ostro2002}) for imaging the surface of the asteroid and/or reconstructing its 3D shape.  For this purpose, signals are sent from a transmitting antenna (transmitting frequency $F_\mathrm{tx}$) to the asteroid and its echo is registered in a receiving station. Range information is obtained by time-modulating the signal via binary phase coding or frequency modulation (e.g. chirp). Thus, for an appropriate SNR the maximum spatial resolution is determined by the \rrv{decoder bandwidth} of the transmitter. The Doppler shifted, echoed signal is analysed in time and frequency domain to construct an image data set containing dynamical and spatial information (for a review of the technique see \citealt{ostro1993,margot2021} and references therein). 

One of the first and foremost opportunities of the \Apophis approach will be to \rvv{constrain} its orbit. Being a Potentially Hazardous Asteroid (PHA), all the information that we can obtain to further \rvv{constrain} the non-gravitational forces acting on the asteroid, is required to predict future approaches and assess the probability of an impact.

\begin{figure}
    \centering
    \includegraphics[width=\linewidth]{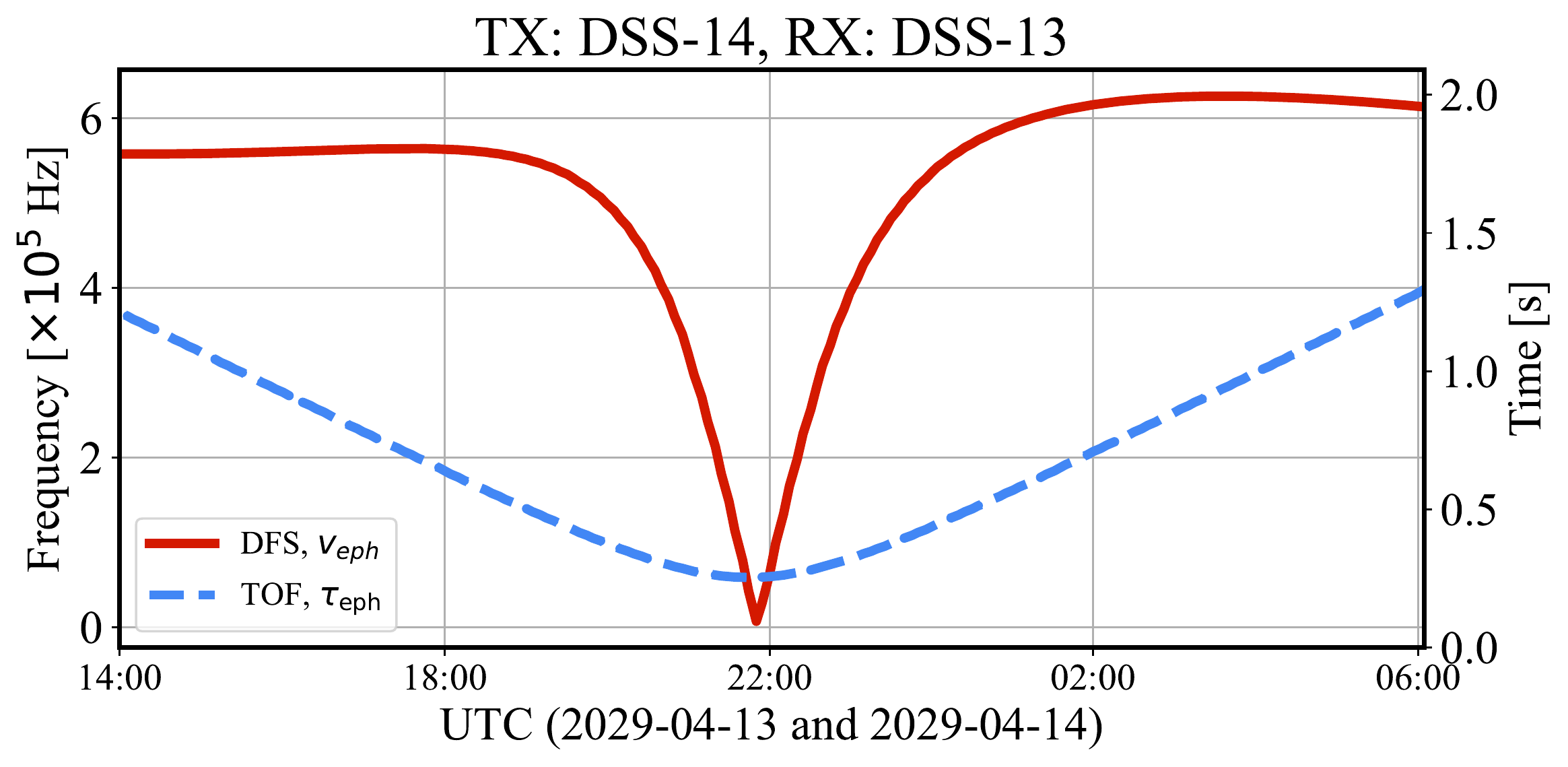}
    \caption{\rev{Round-trip time} $\tau_\mathrm{eph}(t)$ and center-of-mass Doppler frequency shift $\nu_\mathrm{eph}$, for the radar signal in the case of bistatic radar observations using the antennas DSS 14 \rvv{and} DSS-13 of the Goldstone Complex.}
    \label{fig:TOF_DFS}
\end{figure}

In order to achieve this, radar observations measure the round-trip time (the delay) of the signal $\tau_\mathrm{eph}(t) = 2d(t)/c$, \rvv{where $d(t)$ is} the instantaneous distance between the target and the radar, and the Doppler frequency shift (DFS), $\nu_\mathrm{eph}(t)=2F_\mathrm{tx} v(t)/c$ of the center-of-mass of the asteroid, where $v(t)$ is the instantaneous target-radar relative radial velocity.  In \autoref{fig:TOF_DFS} we show the typical value of these two critical quantities for observations during the approaching window using the DSS-14 and DSS-13 antennas at the Goldstone complex.

The expected range resolution for each observation depends on the sampling capabilities of the transmitting radar system. For instance, when using the pulse compression technique \citep{Naidu_2013} the range resolution $\delta r$ is calculated from the decoder effective bandwidth $B_\mathrm{eff}$ as follows,
\begin{equation}
    \delta r = \frac{c}{2}\frac{1}{B_\mathrm{eff}}\ .
\end{equation}

The echoed signal will produce a \rrr{delay dispersion of $\sim D_{||}/(2\delta r)$, where $D_{||}$ is the spatial ``length'' of the asteroid in a direction parallel to the viewing line} (which for the case of Apophis would be in the 300-370 m range depending on its relative orientation), and a Doppler shift dispersion (or bandwidth),

\begin{equation}
\Delta \nu=\frac{2D\omega\cos\delta}{\lambda}\ .
\label{eq:delta-nu}
\end{equation}

\rrr{Here, D is the spatial ``width'' of the asteroid, measured orthogonal to the line of sight (since \apophis is not spherical, in general we can expect $D_{||}$ to be different to $D$).} $\delta$ is the target subradar latitude and $\omega$ is the apparent angular frequency of the asteroid which will change during the passage due to the combined effects of its intrinsic rotation and the apparent rotation due to the change in viewing perspective as it approaches Earth (see below). The subradar latitude $\delta$ will also change during the passage due to the asteroid orbit not being coplanar with the Earth's equator. This means that the Doppler bandwidth will not remain constant during the passage. \rrv{Here we consistently update this value for our per-run SNR calculation (see \autoref{eq:trun})}. To sample this bandwidth with an adequate frequency resolution, the integration time per run needs to be at least,
\begin{equation}
    \Delta t_\text{run} = \Delta \nu^{-1}\ .
\label{eq:trun}
\end{equation}

For monostatic observations, the TX-RX switch time limits the highest possible frequency resolution to $(\tau_\mathrm{eph}-t_\text{switch})^{-1}$. This restriction does not apply to bistatic observations, and the integration time is only limited by the time the asteroid can be tracked by the telescope pair, i.e. one observation track.

The achievable SNR will depend on many factors, involving the TX/RX instrumentation as well as the target properties.  \rvv{For instance, one of the main quantities is the total received echo power, which depends on the antenna gains $G$, both for the transmitting} and receiving antennas \citep{Naidu_2016},

\begin{equation}
G=4\pi \eta A/\lambda^2\ .
\end{equation}
Here $\eta$ is the aperture efficiency, $A$ is the \rvv{physical area of the antenna}, and $\lambda$ is the radar operating wavelength (which depends on the configuration of the \rvv{transmitting} antenna). The aperture efficiency for most paraboloidal antennas considered here (see Table~\ref{tab:observatories}) is in the 0.6-0.7 range \citep{Naidu_2016}, \rvv{although FAST is on the lower end of this range} \citep{dong_han_2013}. Once the transmitted power $P_\mathrm{tx}$ is set, the received power $P_\mathrm{rx}$ can then be calculated as \citep{Naidu_2013},

\begin{equation}
P_\mathrm{rx}=\frac{G_\mathrm{rx}G_\mathrm{tx}\lambda^2 \sigma}{(4\pi)^3R_\mathrm{rx}^2R_\mathrm{tx}^2 }\,P_\mathrm{tx}\ .
\label{eq:Prx_Ptx}
\end{equation}
Here $\sigma$ is the radar cross-section of the target object, which has been measured as $\sigma\approx{0.023}$ km$^2$ in previous observations \citep{Brozovic2018a},  and $R_\mathrm{tx}$, $R_\mathrm{tx}$ are the distances from the TX and RX antennas to the target, which are usually assumed to be approximately the same, e.g. $R_\mathrm{rx}\approx R_\mathrm{tx}$ \citep{Naidu_2016}. In the case of the 2029 \Apophis passage, however, the short distance range involved does not allow for this approximation, and a proper calculation of both quantities is required. \rvv{It should be noted that the transmitted peak power for pulsed radar systems such as HUSIR and TIRA is not the same as the effective average power, as discussed above}.



\rvv{The SNR is the echo-power standard $z$-score of the signal, i.e. the number of standard deviations above the mean noise power},
\begin{equation}
\mathrm{SNR}=\frac{P_\mathrm{rx}}{\Delta P_\mathrm{rms}}\ .
\label{eq:SNR}
\end{equation}
Here $\Delta P_\mathrm{rms}$ is the noise \rvv{standard deviation} which depends on the system temperature of the receiving telescope $T_\mathrm{sys}$, the Doppler bandwidth $\Delta \nu$, and the integration time $\Delta t$,

\begin{equation}
\Delta P_\mathrm{rms}=kT_\mathrm{sys}\sqrt{\frac{\Delta \nu}{\Delta t}}\ .
\label{eq:Prms}
\end{equation}
Here $k$ is the Boltzmann constant. 

Replacing Eqs. (\ref{eq:delta-nu}), (\ref{eq:Prx_Ptx}) and (\ref{eq:Prms}) into Eq. (\ref{eq:SNR}) the final expression for the signal-to-noise ratio of bistatic radar observations becomes,

\begin{equation}
\mathrm{SNR}=P_\mathrm{tx}\frac{G_\mathrm{rx}G_\mathrm{tx}\lambda^{5/2}}{(4\pi)^3 kT_\mathrm{sys}}\sqrt{\frac{\Delta t}{2D\omega\cos\delta}} \frac{ \sigma}{R_\mathrm{rx}^2R_\mathrm{tx}^2 } \ .
\label{eq:SNR_full}
\end{equation}

\begin{figure}
    \centering
    \includegraphics[width=\linewidth]{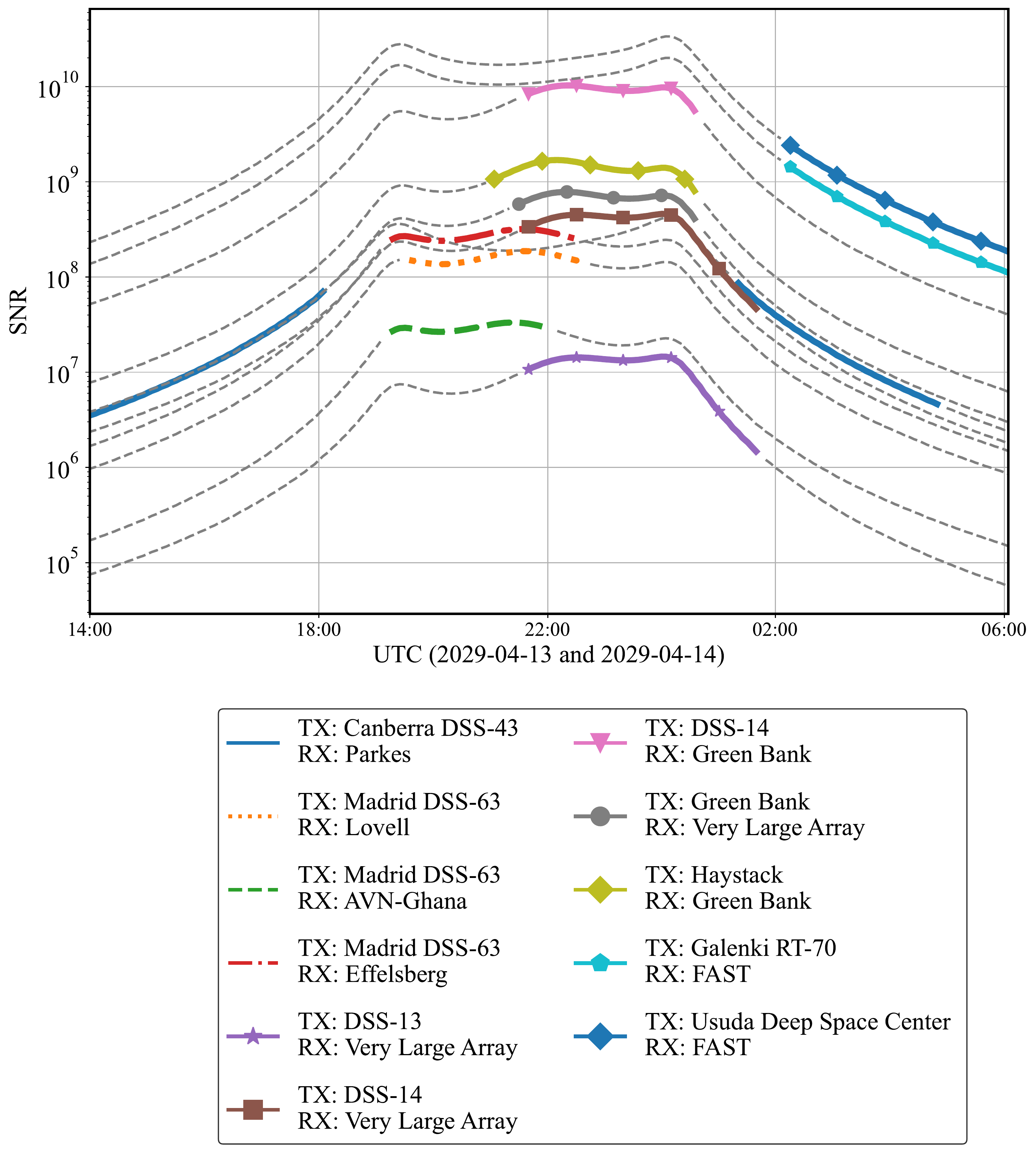}
    \caption{Expected values of the frequency integrated SNR per run for different pairs of radio observatories during the 2029 \Apophis approaching window. The grey dotted lines correspond to times in which Apophis is under the minimum elevation for either of the paired antennas. \rrr{The run time is calculated here according to \autoref{eq:trun}.}
}
    \label{fig:SNR}
\end{figure}

In \autoref{fig:SNR} we show our simulated values of the expected \rvv{frequency integrated} SNR per run \rrv{(see \autoref{eq:trun})} for bistatic radar observations of the 2029 \Apophis approach for different observatory pairs and \rvv{calculated} with \autoref{eq:SNR_full} and the antenna properties listed in \autoref{tab:observatories}. We assume \rvv{a} nominal system temperature of 25 K. This temperature may be in some cases higher by a factor \rrr{of 2, or more when specified,} which means that our calculated SNR may be off by \rrr{the same factor}. We do not consider this to be a significant problem since the lowest estimated SNR is \rvv{in} the order of $10^7$, for the case of the DSS-13 + VLA pairing. Even in the lowest SNR case (observations performed between DSS-13 and a VLA element), values of $10^4$ can be achievable, which is well suited for radar observations. With Arecibo out of operation, observations with FAST + Usuda will provide the highest SNR per run during the whole set of observations.

\section{Rotation, axis orientation and coverage} 
\label{sec:rotation}

\begin{figure}
    \centering
    \includegraphics[width=\linewidth]{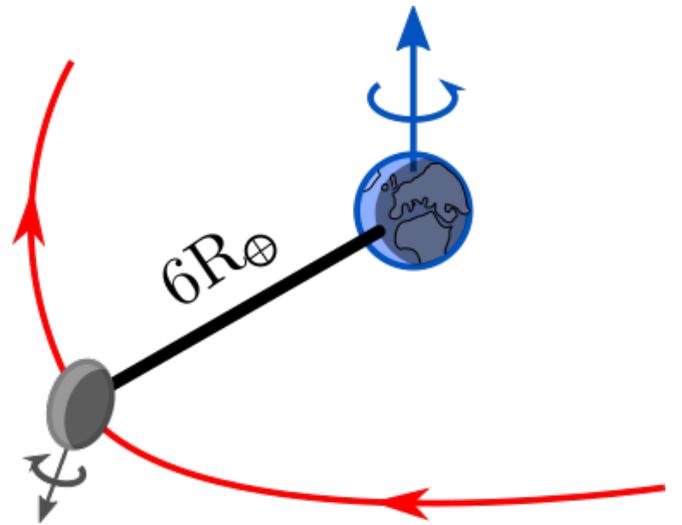}
    \caption{Schematic illustration of the trajectory of Apophis during the 2029 approach, showing the most probable orientation of its spin axis. For illustration purposes, \Apophis size has been increased by a factor of 30 000, and its geometrical main axis pointing in the same direction as the rotational, although there is no strong evidence of the long principal axis being aligned with the spin axis. The asteroid main axis of rotation is pointing at roughly the southern celestial pole (J2000 RA,DEC = $118.8^{\circ},-79.4^{\circ}$).}
    \label{fig:orbit-rotation}
\end{figure}

Using previous observations of \Apophis and theoretical models, several authors have been able to constrain the rotational state of the asteroid \citep{Brozovic2018a}. It is now known that the spin axis of \Apophis (which is also the longest one) probably points towards RA $118.8^{\circ}$ and DEC $-79.4^{\circ}$ in J2000 coordinates \citep{Pravec2014,Souchay2018}. Moreover, Apophis seems to be a non-principal axis rotator with a main sidereal rotation period of 30.56 h \citep{Brozovic2018a} which  corresponds to an intrinsic angular speed $\omega_\mathrm{s}=11.7$ deg/h.  In \autoref{fig:orbit-rotation} we schematically show the expected orientation of Apophis spin during its 2029 approach.

In order to take into account the expected spin state on the observations (delay-Doppler observations and surface coverage), we need to predict how the surface will be rotating with respect to the radar plane-of-sky (POS) reference frame. \rev{Although its current rotation speed is well constrained, according to \cite{Souchay2018} significant variations in obliquity and precession in longitude may occur due to gravitational torques during the close approach. However, the single most significant effect altering the observed rotational speed will be the change in the Earth-facing side of the asteroid due to the change in perspective as the asteroid swings by. Thus, the swiftness of the asteroid approach will temporarily increase its apparent rotational speed as seen from Earth, which will increase the surface coverage due to the change in visual perspective.} The apparent angular speed of the asteroid in the POS can be obtained from the vector sum \citep{ostro2002},
$$
\vec{\omega}_\mathrm{app}=\vec{\omega}_\mathrm{s}+\vec{\omega}_\mathrm{sky}\ .
$$
Here $\vec{\omega}_\mathrm{s}$ is the sidereal (intrinsic) spin-vector and $\vec{\omega}_\mathrm{sky}$ is the \rev{sky angular speed} which is given by,

$$
\vec{\omega}_\mathrm{sky}=\dot{\hat e}\times {\hat e}\ .
$$

Here, $\hat e=\vec r/r$ is the target-to-radar unit vector, and $\dot{\hat e}$ is its vectorial speed that can be calculated from ephemeris using,

$$
\dot{\hat e}=\frac{\vec v}{r}-\frac{\vec r}{r^2}\dot r\ .
$$

\begin{figure}
    \centering
    \includegraphics[width=\linewidth]{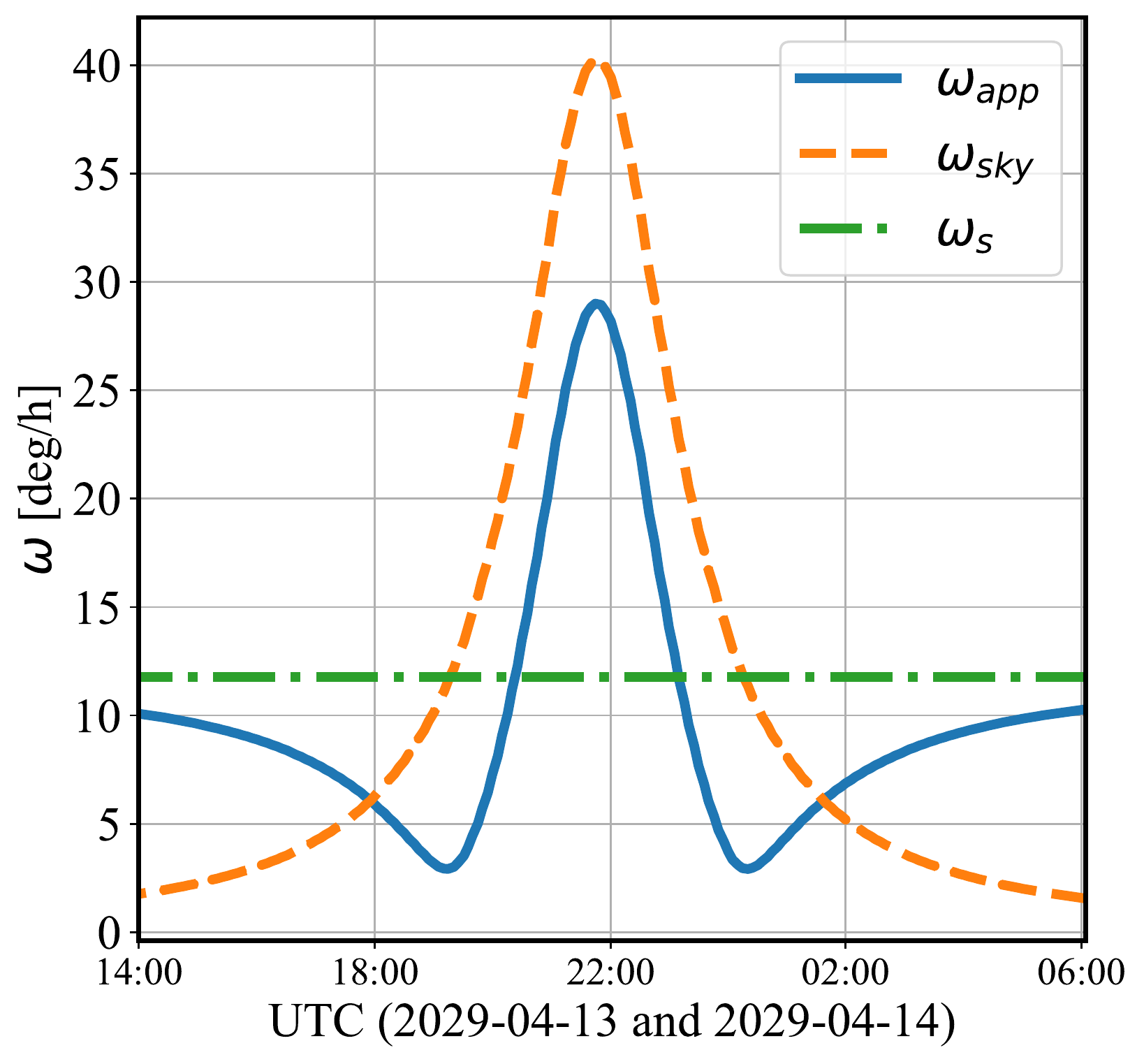}
    \caption{Apparent rotational speed of the asteroid in the hours around the closest approach (solid line). The rotational speed of the asteroid is a combination of its intrinsic rotation ($\omega_\mathrm{s}$) and the sky angular speed of the asteroid during its closest approach ($\omega_\mathrm{sky}$).}
    \label{fig:apparent-rotation}
\end{figure}

In \autoref{fig:apparent-rotation} we show the value of $\vec{\omega}_\mathrm{sky}$ and  $\vec{\omega}_\mathrm{app}$, as a function of time for the hypothetical case of a radar in the center of the Earth.

As expected, given the extremely close and rapid approach of \Apophis, $\omega_\mathrm{sky}$ will reach values ${\sim}4$ times larger than the intrinsic rotational velocity of the asteroid.  However, since the spin axis of the asteroid is almost orthogonal to its orbital plane, the apparent rotation will possibly compensate its intrinsic rotation.  As a result, \rvv{during 10 of the 14 hours} of the approaching window the asteroid will likely rotate slower than usual. At certain times, its surface will even seem static with respect to Earth (minimum values in the \rvv{solid blue line} in \autoref{fig:apparent-rotation}).  

However, during the closest approach (between UTC 20:00 and 23:59 of April 13) the sky rotation is higher than its intrinsic rotation and the asteroid will seem to be rotating ${\sim}2-3$ times faster.  This will increase the Doppler shift dispersion (see \autoref{eq:delta-nu}).  More interestingly, this boost in the apparent rotation will allow observations to cover a large \rrv{fraction of the asteroid surface at potentially better range resolutions than in observing campaigns previous to this approach}.
 
If we manage to perform the coordinated radar observations program proposed in this paper, an interesting question to raise is what fraction of the asteroid surface will be covered by observations.  Since the approaching window  (${\sim}14$ hours) is almost half of a full rotation of the asteroid (${\sim}30$ hours) and the apparent rotation happens in the opposite direction (see above), it would be expected that not all the surface will be imaged at the \rvv{plausible resolution shown} here. 

\begin{figure*}
    \centering
    \includegraphics[width=\linewidth]{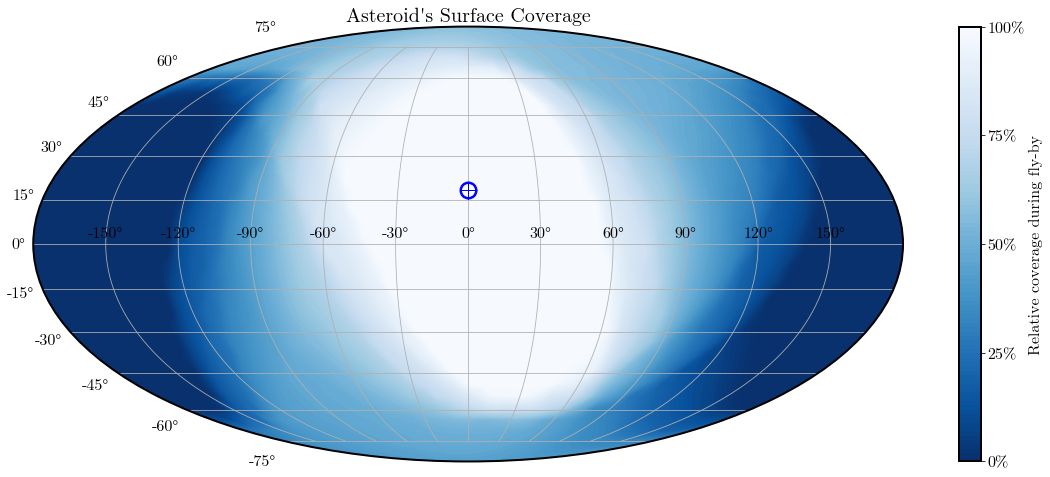}
    \caption{Surface coverage of \Apophis assuming a continuous observation during the 14 hours around the closest approach. The reference meridian (lon$_A$=0) is defined by the sub-Earth direction at the time of closest approach. A value of 50\% over a point in the map means that this point can be observed during half of the approaching window.}
    \label{fig:coverage}
\end{figure*}

In order to predict the radar coverage of the asteroid's surface, we simulated the Earth's apparent sky motion as seen from many points on Apophis' surface.  For this purpose, we used a spherical model of the asteroid, create a grid of local coordinates, and \rvv{calculate} for each point and at every time during the approaching window, if the geocenter was above the local horizon.  In \autoref{fig:coverage} we show the resulting coverage map.

The Apophis geographical coordinates were defined with respect to its axis of rotation (defining the latitude) and the direction of Earth in the moment of closest approach (defining longitude).  The prime meridian in this system pass through the sub-Earth point on the surface of the asteroid at the time of closest approach, namely 2029-04-13 at 21:45 UTC.  

Due to the inclination of the asteroid rotation axis, the coordinates of the sub Earth point at that time will be lon$_{A}=0^\circ$, lat$_{A}=26.4^\circ$ (white circle in \autoref{fig:coverage}). All the radio observatories included in this work will be projected within an area of $10^\circ$ around this point and hence they will cover approximately the same area of the asteroid, if visible from their location. 


It should be noted that Apophis will not do a full revolution in the time window described in this paper, and in the best case scenario 85\% of the surface will be partially observable during the 14 hour \rvv{window of interest}. The coverage of this missing zone will have to rely on observations taken in the months \rvv{before and after the closest approach, during which a high SNR will be achievable for many of the radar pairs suggested here, although possibly not at the highest possible range resolution}.

Finally, we also note that Apophis may experience a spin rate change during its close approach due to tidal effects. Thus, the actual intrinsic spin axis of Apophis may be different than we assume here. However, given that the main source of apparent rotation is the change of perspective due to the sky speed of the asteroid during its rapid orbital approach, the apparent spin axis \rvv{will not be too different from our theoretical estimations shown here}.

\section{\rvv{Summary} and Discussion}

The 300 m asteroid Apophis will fly within six earth radii from geocenter on April 13th 2029. Although \rvv{there is no collision risk, the possibility is not yet ruled out for future encounters. Further studies of the asteroid surface, shape and orbital changes have yet to be made, and it is during this close encounter when the scientific opportunities to study this object are the greatest. We present in this paper a sequence of possible radar observations of Apophis within a 14 hr window near} the closest approach and evaluate which radio facilities will be optimal for complementary observations of the asteroid. Given that the next windows of opportunity for radar ranging during a close passage of Apophis will be in 2044 and 2051 and at much larger distances \citep{sokolov}, we should take full advantage of this opportunity.

We use trajectory and ephemeris data provided by the JPL Horizons Web Service to visualize the path of Apophis around the world (\autoref{sec:approach}). We show that it will fly over some of the world's most important radio telescopes, such as Parkes, GBT, FAST, among others, as well as over antennas with powerful \rvv{transmitting} capabilities: Canberra DSS-43, Madrid DSS-63 \citep{dsn70m}, \rev{HUSIR and Goldstone DSS-13, DSS-14}, among others. \rvv{Based on this information we propose a 4-stage observation plan using TX/RX capable antenna pairs under the asteroid path in order to obtain continuous and overlapping information of this object}. 

The 4-stage observation plan will help scheduling radar observations of the asteroid in the near future, as well as coordinating scientific efforts in countries under the flight path. For example: In stages 1 and 2 (Australia and Africa/Europe) Apophis will be visible even with the naked eye in the night sky. During stages 2 and 3 (Africa/Europe and USA) the apparent rotation of Apophis will increase 2.5x, and \rvv{the per frequency bin SNR will correspondingly increase with respect to similar pairings in other stages}. As for stage 4 (east Asia) the asteroid is going to fly over the Chinese 500 m dish (FAST), \rvv{which can be paired with the TX capable} deep space antennas in Usuda, Japan, or in Galenki (Ussuriysk), Russia \rvv{if pairing frequency issues are resolved}.

Following this plan, the Goldstone DSS-14 + Green Bank pairing will yield the highest possible \rrv{per run}, frequency integrated SNR ($\sim10^{10}$), but the DSS-13 + VLA and HUSIR + GBT pairings will yield the best range resolutions, reaching near-cm/px at \rrv{per run}, integrated SNRs above $\sim10^{5-6}$ (Figure~\ref{fig:SNR}). We have found suitable pairings around the world that ensure that bistatic radar pairs are always mapping the asteroid during its close approach, \rvv{thus achieving an unprecedented multi-wavelength coverage of the asteroid. This} will drastically improve our knowledge of its astrometry, shape, composition, and regolith size distribution.

We have shown that powerful antennas are necessary for optimal bistatic observations of Apophis. With this in mind, we strongly advise in favour of the reconstruction of a next generation-Arecibo Radio Telescope such as the 1 MW-transmitter instrument proposed by \citet{roshi2021}, as its geographical location and \rvv{transmitting} power would make it the optimal transmission facility for this task, and the data gathered by the US bistatic radar pairs would be the best we could get out of this asteroid's close flight.

\rvv{We also encourage the use of the} coordinate system proposed in \autoref{sec:rotation}. This would facilitate joint efforts in describing and modeling the shape of Apophis. 

Finally, we consider the fly-by of Apophis as a perfect opportunity to engage the general public with science, and we advise the scientific community and scientific journalists to start planning for this date. In countries around the stages 1 and 2 (see \autoref{sec:coordinated}) everyone can experience the naked-eye sight of a 350 m asteroid, which will offer unprecedented astrophotography opportunities of an asteroid. 

In countries where daylight obstructs the view, citizenship science projects can be proposed based on the way bistatic radar works: If a powerful enough antenna is bouncing waves off Apophis, even small antennas might be able to detect signals coming off it. This is because the expected SNRs will be so high (our lowest SNR for DSS-13 + VLA is in the order $10^7$) that even with a 1-m size antenna an SNR (per 2 minutes integration time) \rvv{above $10^2$ could be expected, assuming a $T_\text{sys}\sim300$ K}. This also applies for other bistatic radar pairs in regions of the world. And although it is unlikely that amateur antennas will have powerful enough decoders to \rvv{delay-Doppler resolve Apophis, the SNR may be just high enough so amateur radio astronomers who prepare their instrumentation for pair-specific frequencies may be able to pick up bounce signals}.

\section{Data Availability}

All the computations provided in this paper can be reproduced using the scripts and tools available in the {\tt GitHub} repo \url{https://github.com/seap-udea/Apophis2029}. They can also be executed as a {\tt Google Colaboratory} using the link \url{https://bit.ly/Apophis2029-Colab2}. 

\section{Acknowledgements}

Most of the computations that made possible this work were performed with Astropy \citep{astropy:2018}, Cartopy \citep{Cartopy}, and their related tools and libraries, {\tt iPython} \citep{Perez2007}, {\tt Jupyter} \citep{Perez2007}, {\tt Matplotlib} \citep{Hunter2007}, {\tt scipy} and {\tt numpy} \citep{Van2011}. The authors are grateful with the anonymous referee, whose comments and suggestions significantly improved this work. \rev{We especially appreciate the useful insights and and numerous comments of Dr. Joseph Lazio}. 

\bibliographystyle{mnras}
\bibliography{manuscript} 

\bsp
\label{lastpage}
\end{document}